\documentclass[useAMS, usenatbib, usegraphicx, referee]{biom}
\usepackage{amssymb}
\usepackage{mathtools}
\usepackage{graphicx}
\usepackage{booktabs, caption}
\usepackage{subcaption}
\usepackage [section]{placeins}
\usepackage{mathrsfs}
\usepackage{bbm}
\newtheorem{proposition}{Proposition}
\usepackage [english]{babel}
\usepackage [autostyle, english = american]{csquotes}
\MakeOuterQuote{"}
\DeclareGraphicsExtensions{.pdf,.png}
\title[Counterfactual fairness for small subgroups]{Counterfactual fairness for small subgroups}

\author{Solvejg Wastvedt$^*$\email{wastv004@umn.edu}, 
Jared D Huling, and 
Julian Wolfson \\
Division of Biostatistics, University of Minnesota, Minneapolis, Minnesota, U.S.A.}

\begin{document}





\pagerange{\pageref{firstpage}--\pageref{lastpage}} 




\label{firstpage}


\begin{abstract}
While methods for measuring and correcting differential performance in risk prediction models have proliferated in recent years, most existing techniques can only be used to assess fairness across relatively large subgroups. The purpose of algorithmic fairness efforts is often to redress discrimination against groups that are both marginalized and small, so this sample size limitation often prevents existing techniques from accomplishing their main aim. We take a three-pronged approach to address the problem of quantifying fairness with small subgroups. First, we propose new estimands built on the "counterfactual fairness" framework that leverage information across groups. Second, we estimate these quantities using a larger volume of data than existing techniques. Finally, we propose a novel data borrowing approach to incorporate "external data" that lacks outcomes and predictions but contains covariate and group membership information. This less stringent requirement on the external data allows for more possibilities for external data sources. We demonstrate practical application of our estimators to a risk prediction model used by a major Midwestern health system during the COVID-19 pandemic.
\end{abstract}

%



\maketitle


%

\section{Introduction} \label{sec:introduction}

Increasing use of complex risk prediction models in health care settings has drawn attention to both the opportunities and the challenges such models present. Clinical risk prediction models can improve care for patients and efficiency for providers by personalizing treatment, identifying high-risk patients for early intervention, and more. However, the models, which often use opaque techniques that are not readily understandable by patients and providers, have also demonstrated the potential to create and entrench health inequities. One well-documented strand of this problem is the potential for risk prediction models to perform more poorly for some groups than others. In clinical settings, inaccurate model predictions can lead to sub-optimal assignment of treatment and worse health outcomes for the affected groups, which are often those already marginalized in society \citep{obermeyerDissectingRacialBias2019a}.

While techniques for assessing and correcting model bias have proliferated in recent years (e.g., \citet{castelnovoClarificationNuancesFairness2022a}, \citet{chenComprehensiveEmpiricalStudy2023}), much of this work does not address a major challenge in the clinical setting: limited sample size in the smallest groups where model performance is to be assessed to ascertain fairness. As an example, our previous work, which proposes estimators of intersectional unfairness adapted for clinical situations in which a treatment is in use, requires a reasonably large sample size in all groups to obtain useful precision \citep{wastvedtIntersectionalFrameworkCounterfactual2023}. In that work, we analyzed a risk prediction model used during the COVID-19 pandemic to help determine whether a patient should be transferred to a COVID-19 cohort hospital. We looked at unfairness across intersecting race and age categories, but we were only able to separate race into white, Black or African American, and a group for all other races because of limited sample size. While this analysis reveals differing model error rates by age and for patients identifying as white vs. Black or African American, the "all other races groups" has little utility. The experiences of individuals in this group, which includes for example both Asian and American Indian patients, differ widely. It is likely that the model performs differently for the subgroups encompassed in this category, but those error rate differences disappear when the subgroups are examined together. More broadly, lumping small subgroups together in this fashion can erase the experiences of these patients from analyses. The main purpose of fairness assessment is to uncover mistreatment of marginalized groups which are often, though not always, numeric minorities as well. Thus it is crucial to the goals of fairness methods that they prioritize the ability to obtain results for small subgroups.

In this work, we address the problem of limited sample size by proposing new estimators that are estimable and have reasonable precision even for very small groups. We extend the COVID-19 model application presented in our earlier work to show the utility of these new estimators in situations where existing methods fail. The COVID-19 risk prediction model we analyze is one of many that were used to help guide allocation of resources in high-pressure scenarios as case numbers rose. The model we analyze was trained by the health system on 1,469 adults who tested positive for SARS-CoV-2. Features used for prediction included patient demographic variables, home medications, and medical conditions. Data available for our after-the-fact assessment of this model is limited because of the ever-changing nature of COVID-19 treatment and the relatively short period, during the height of the pandemic, over which the model was used. As is common in clinical risk prediction scenarios, the population for which the model was used is small in comparison to the population served by the health system. Thus there is potential to borrow information from other available data drawn from the same population in which risk scores and outcomes are not recorded; we explore this possibility in our proposed methods.

In the sections that follow we first describe the \emph{counterfactual error rates}, model performance metrics developed in existing work \citep*{costonCounterfactualRiskAssessments2020, mishlerFairnessRiskAssessment2021a} that are well-suited to clinical applications. We then take a three-pronged approach to improving these metrics' performance for small subgroups. First, we re-formulate the error rates to leverage information across groups by incorporating the overall error rate. We then use the causal identification process to arrive at estimators that use more of the data than existing methods. Finally, we propose a novel data borrowing procedure that uses "external data", such as that available in our COVID-19 risk model application, to further improve estimation.

\section{Statistical framework} \label{sec:framework}

Following algorithmic fairness convention, define a \emph{protected characteristic} as any grouping, such as race or gender, along which we wish to measure discrimination. Let $\pmb{A}$ denote a vector of categorical protected characteristics, in which each element indicates group membership for one characteristic. Let $m$ be the number of protected characteristics that we wish to consider. Denote the characteristics $A_j$, $j \in \{1, ..., m\}$ and assume each $A_j$ is a categorical variable with a finite number of levels, the set of which is denoted $\mathcal{A}_j$. Let $\pmb{A} = \{A_1, A_2, ..., A_m\}^T \in \mathcal{A}$ contain all protected characteristics of interest, where $\mathcal{A}$ is the set of all possible combinations of all levels of the $m$ characteristics.

To finish notation, let $S$ denote a binary risk prediction. Although many clinical risk models produce predicted probabilities, we are using the binary prediction derived by selecting a cut-off threshold. Let $D$ denote a binary treatment assignment and $Y$ a binary outcome such as an adverse health event. Under a binary treatment, there are two potential outcomes: $Y^0$, the outcome under no treatment, and $Y^1$, the outcome under treatment. We focus on the $Y^0$ outcome as it represents patients' baseline risk and thereby is informative for guiding treatment.

Existing work \citep{mishlerFairnessRiskAssessment2021a} defines the following counterfactual versions of the common false positive rate and false negative rate model performance metrics for a single protected characteristic. In our previous work, we apply the definitions to the vector of protected characteristics $\pmb{A} = \pmb{a}$.

\begin{definition} \label{def:cferr}
The \emph{counterfactual false positive rate} of a predictor $S$ for the group having protected characteristic vector $\pmb{A}=\pmb{a}$, denoted $cFPR(S,\pmb{a})$, is equal to $Pr(S=1 | Y^0 = 0, \pmb{A} = \pmb{a})$. The \emph{counterfactual false negative rate}, $cFNR(S,\pmb{a})$, is equal to $Pr(S=0|Y^0 = 1, \pmb{A} = \pmb{a})$. 
\end{definition}

Our previous work proposed the following weighted estimators of $cFPR(S,\pmb{a})$ and $cFNR(S,\pmb{a})$:

\begin{align}
    \widehat{cFPR}(S,\pmb{a}) &= \frac{\sum_{i=1}^n[(1-D_i)S_i(1-Y_i) \mathbbm{1}(\pmb{A}_i=\pmb{a})/(1-\hat{\pi}_i)]}{\sum_{i=1}^n[(1-D_i) (1-Y_i)\mathbbm{1}(\pmb{A}_i=\pmb{a})/(1-\hat{\pi}_i)]} \label{cfpr_est} \\
    & \nonumber \\
    \widehat{cFNR}(S,\pmb{a}) &= \frac{\sum_{i=1}^n[(1-D_i)(1-S_i)Y_i \mathbbm{1}(\pmb{A}_i=\pmb{a})/(1-\hat{\pi}_i)]}{\sum_{i=1}^n[(1-D_i)Y_i \mathbbm{1}(\pmb{A}_i=\pmb{a})/(1-\hat{\pi}_i)]} \label{cfnr_est}
\end{align}.

The \emph{counterfactual error rate differences}, denoted $\Delta^+(S)$ and $\Delta^-(S)$, quantify the differences in counterfactual false positive and false negative rates between two protected characteristic groups.

\section{New estimands to borrow strength across groups} \label{sec:estimands}

The estimators for $cFPR(S,\pmb{a})$ and $cFNR(S,\pmb{a})$ in equations \eqref{cfpr_est} and \eqref{cfnr_est} present a challenge in practical applications with limited sample size. The estimators are weighted averages over a subset of a subgroup defined by intersecting protected characteristics. In the case of $cFPR(S,\pmb{a})$, only observations with $D_i = 0$, $S_i = 1$, and $Y_i = 0$ contribute non-zero components to the average. The estimator for $cFNR(S,\pmb{a})$ is analogous. Restricting the estimation to this slice of what may already be a small protected subgroup can make the estimators unstable.

To address this issue, we propose alternative estimators for the counterfactual error rates that incorporate more of the data. Our three-pronged approach begins by reformulating the estimands for group-specific error rates to leverage the overall error rate. Next, we make an additional assumption during causal identification to reduce reliance on indicator variables. Finally, the next section explains potential use of an auxiliary data set during estimation to further improve performance.

Our re-formulation of the counterfactual error rates draws on \citet{efronLargeScaleInferenceEmpirical2010}, who propose rewriting the false discovery rate (FDR) of a predictor $S$ for a particular subgroup as a ratio of probabilities multiplied by the overall FDR. Following a similar process, we obtain alternate expressions for the counterfactual error rates given in Proposition \ref{prop:rewrite} (proof in Web Appendix A).

\begin{proposition} \label{prop:rewrite}
Given two binary variables $S$ and $Y^0$ and a subgroup vector $\pmb{A} = \pmb{a}$, the counterfactual false negative rate and counterfactual false positive rate of $S$ in group $\pmb{a}$ can be rewritten as the following:

\begin{align}
    cFPR(S, \pmb{a}) &= cFPR(S)\frac{P(\pmb{A}=\pmb{a} | Y^0 = 0, S = 1)}{P(\pmb{A}=\pmb{a} | Y^0 = 0)} \label{eq:cfpr_rewrite}\\
    cFNR(S, \pmb{a}) &= cFNR(S)\frac{P(\pmb{A}=\pmb{a} | Y^0 = 1, S = 0)}{P(\pmb{A}=\pmb{a} | Y^0 = 1)} \label{eq:cfnr_rewrite}
\end{align}
\end{proposition}.

We then identify our estimands from observed data using standard causal inference assumptions plus one additional assumption that allows us to avoid many of the indicator variables used in \citet{wastvedtIntersectionalFrameworkCounterfactual2023}, thereby employing a larger volume of data. We assume $Y^0$ is independent of $\pmb{A}$ given $X$, or that we have collected sufficient covariates such that protected group membership gives no additional information about the probability of $Y^0$. A full list of assumptions and simulations demonstrating that our methods are robust to violations of the additional assumption are in Web Appendices B and D.

Using these assumptions, we break down the ratio of conditional probabilities in each error rate expression in Proposition \ref{prop:id}  (proof in Web Appendix B). Let $X$ be a vector of observed covariates and define
the propensity score function for a given protected group as $\pi = P(D = 1|A, X, S)$.

\begin{proposition} \label{prop:id}
The conditional probability ratios in expressions \eqref{eq:cfpr_rewrite} and \eqref{eq:cfnr_rewrite} are identified as follows for $Y^0=y \in \{0,1\}$ and $S=s \in \{0,1\}$:

\begin{equation*}
    \frac{P(\pmb{A}=\pmb{a} | Y^0 = y, S = s)}{P(\pmb{A}=\pmb{a} | Y^0 = y)} = \frac{ E[ \mu_0(y,s,X) \mathbbm{1}(\pmb{A}=\pmb{a}) \mathbbm{1}(S=s) ] / E[ \mu_0(y,s,X) \mathbbm{1}(S=s) ] }{ E[ \mu_0^*(y,X) h(X,\pmb{a}) ] / E[ \mu_0^*(y,X) ] } \text{,}
\end{equation*}

where $\mu_0(y,s,X) = P(Y=y|D=0,S=s,X)$ and $\mu_0^*(y,X) = P(Y=y|D=0,X)$. The $h$ function is defined as $h(x, \pmb{a}) = P(\pmb{A} = \pmb{a}|X=x)$.
\end{proposition}

While the estimators in \citet{wastvedtIntersectionalFrameworkCounterfactual2023} limited data to specific intersections of $D$, $S$, $Y$, and $\pmb{A}$, Proposition \ref{prop:rewrite} allows us to estimate the denominator using all the data and only restrict by the protected characteristic vector ($\pmb{A}$) and model prediction ($S$) in the numerator. Drawing from more data reduces our estimators' variance and enables estimation even in very small subgroups.

The overall counterfactual error rate components of equations \eqref{eq:cfpr_rewrite} and \eqref{eq:cfnr_rewrite} use the identification result in \citet{wastvedtIntersectionalFrameworkCounterfactual2023}, generalized to include all observations: $cFPR(S) = \frac{E [ (1-D)S(1-Y) / (1-\pi(\pmb{A},X,S)) ]}{E [ (1-D)(1-Y) / (1-\pi(\pmb{A},X,S)) ]}$; $cFNR(S) = \frac{E [ (1-D)(1-S)Y / (1-\pi(\pmb{A},X,S)) ]}{E [ (1-D)Y / (1-\pi(\pmb{A},X,S)) ]}$.

\section{Estimation and data borrowing} \label{sec:estimation}

In this section, we demonstrate use of an auxiliary, external data set to aid in estimation of $h(X, \pmb{a}) = P(\pmb{A} = \pmb{a}|X)$. This procedure can reduce variance and bias compared to estimation using only the main data. We then propose regression estimators for the remaining components of the probability ratios in Proposition \ref{prop:id}. 

Throughout, let $\{\pmb{A}_i, D_i, Y_i, X_i, S_i\}$, $i = 1, ..., n$ be the observed data and binary risk predictions. Then the counterfactual error rates for protected group $\pmb{A} = \pmb{a}$ are estimated as follows, where for conciseness we assume that $\hat{\mu}_0$ and $\hat{\mu}_0^*$ are the estimated versions with $y=1$:

\begin{align}
    \widehat{cFPR(S, \pmb{a})} &= \widehat{cFPR(S)} \frac{\sum_{i=1}^n (1-\hat{\mu}_0(s=1,X_i)) \mathbbm{1}(\pmb{A}_i=\pmb{a}) S_i / \sum_{i=1}^n (1-\hat{\mu}_0(s=1,X_i)) S_i }{ \sum_{i=1}^n (1-\hat{\mu}_0^*(X_i))\hat{h}(X_i,\pmb{a}) / \sum_{i=1}^n (1-\hat{\mu}_0^*(X_i)) } \label{eq:cfpr_est} \\
    & \nonumber \\
    \widehat{cFNR(S, \pmb{a})} &= \widehat{cFNR(S)} \frac{\sum_{i=1}^n \hat{\mu}_0(s=0,X_i) \mathbbm{1}(\pmb{A}_i=\pmb{a}) (1-S_i) / \sum_{i=1}^n \hat{\mu}_0(s=0,X_i) (1-S_i)}{\sum_{i=1}^n \hat{\mu}_0^*(X_i)\hat{h}(X_i,\pmb{a}) / \sum_{i=1}^n \hat{\mu}_0^*(X_i) } \label{eq:cfnr_est}
\end{align}.

\subsection{Data borrowing to estimate group membership probabilities}

Estimating the group membership probabilities $\hat{h}(X, \pmb{a})$ involves neither prediction ($S$) nor outcome ($Y$) information. In many clinical risk prediction situations, large volumes of data are available in which the outcome and/or prediction are not present, but a rich set of covariates are. For example, in a health system, this external data could be patient records in the same electronic health record system who were not screened for the adverse event or who did not receive risk predictions. Because of the larger sample size, this external data will typically have better representation from small subgroups. If the distribution of $P(\pmb{A} = \pmb{a} | X)$ in the external data is similar enough to that in the test (or "internal") data, we can use the external data to help estimate group membership probabilities. 

To quantify "similar enough", we employ an adaptive data borrowing method that borrows more information if the distributions are more closely aligned and less (or none) if they differ. Let $\hat{h}_E$ be the internal data group membership probabilities estimated using a model trained on the external data. Let $\hat{h}_I$ be the probabilities fit and predicted using only the internal data. We maximize the predictive performance of the function $\hat{h}^* = \hat{\alpha} \hat{h}_E + (1-\hat{\alpha})\hat{h}_I$ to find the borrowing parameter $\hat{\alpha}$. Various metrics could be used for predictive performance. We use the Brier score because it encompasses both concordance and calibration; however, we achieve similar results in simulations using multi-class area under the ROC curve (AUC). The estimates $\hat{h}^*_i$ are then used in equations \eqref{eq:cfpr_est} and \eqref{eq:cfnr_est}.

\subsection{Estimation of remaining quantities}

The estimators in equations \eqref{eq:cfpr_est} and \eqref{eq:cfnr_est} require several additional parameters. The two outcome models, $\mu_0(X_i, S=1)$ and $\mu^*_0(X_i)$, can be estimated using logistic regression or a more flexible method suitable for binary data. The overall counterfactual error rates are estimated using the weighted estimators proposed in \citet{wastvedtIntersectionalFrameworkCounterfactual2023}: $\widehat{cFPR(S)} = \frac{\sum_{i=1}^n[(1-D_i)S_i(1-Y_i)/(1-\hat{\pi}_i)]}{\sum_{i=1}^n[(1-D_i) (1-Y_i)/(1-\hat{\pi}_i)]}$, $\widehat{cFNR(S)} = \frac{\sum_{i=1}^n[(1-D_i)(1-S_i)Y_i/(1-\hat{\pi}_i)]}{\sum_{i=1}^n[(1-D_i) Y_i/(1-\hat{\pi}_i)]}$. The propensity score, $\pi(\pmb{A},X,S)$, can be estimated using regression methods similar to the outcome models. 

Below we compare estimation of propensity score and outcome models using a generalized linear model and a Super Learner \citep{laanSuperLearner2007}, a more complex ensemble approach that combines multiple machine learning models. Because of the simplicity of the generalized linear models, we use the full sample without any sample splitting or cross-fitting. For the ensemble models, we use a 10-fold cross-fitting approach. For each fold, we fit the three nuisance parameter models on the remaining data to obtain predictions for the held-out fold.

\section{Simulations} \label{sec:sims}

In this section we demonstrate the benefits of our proposed estimators compared to existing methods. We compare to the counterfactual estimators in \citep{wastvedtIntersectionalFrameworkCounterfactual2023} that are not adapted for small subgroups.

All simulations consider two binary protected characteristics, $A_1$ and $A_2$, where for each level $1$ is less common. Thus the group $A_1=0$, $A_2=0$ is the numeric majority, and the group $A_1=1$, $A_2=1$ is the numeric minority. We denote the other groups $M1$ and $M2$ by the protected characteristic ($A_1$ or $A_2$) that is equal to $1$. Overall internal data sample size is limited in all scenarios to create small groups ($N_{internal} = \{50, 100, 150, 200\})$. External data is large ($N_{external} = 10,000$) such that all groups are adequately represented in the external sample. We focus on $cFNR(S,\pmb{a})$ for conciseness; our conclusions also hold for $cFPR(S,\pmb{a})$.

Following the simulation framework in \citet{mishlerFairnessRiskAssessment2021a}, we first generate a set of data ($N = 1,000)$ for training a random forest risk prediction model. We generate a validation data set ($N=50,000$) to determine the true error rates of the risk model. We use the same risk prediction model for all scenarios. For each scenario, we generate internal and external data sets and estimate our unfairness metrics, considering multiclass area under the ROC curve (AUC) and Brier score as performance metrics for data borrowing. In all scenarios, the internal and external $h(X, \pmb{a})$ models are neural networks with a single hidden layer, $100$ units, and a weight decay parameter of $1$ (R package \textit{nnet}). Full data generation details are in Web Appendix C.

\subsection{Variance reduction for small subgroups} \label{sec:sims-part1}

This section compares our proposed and existing estimators under internal sample sizes that are at the lower end of what would typically be encountered in a clinical risk prediction assessment setting ($N_{int} \in \{100,200,500,1000,2000\}$). We use the Brier score for the data borrowing metric. Outcome models (proposed estimators) and the propensity score model (comparison estimators) are correctly specified GLMs. 

\begin{figure}[h]
    \centering
    \includegraphics[width=.8\textwidth]{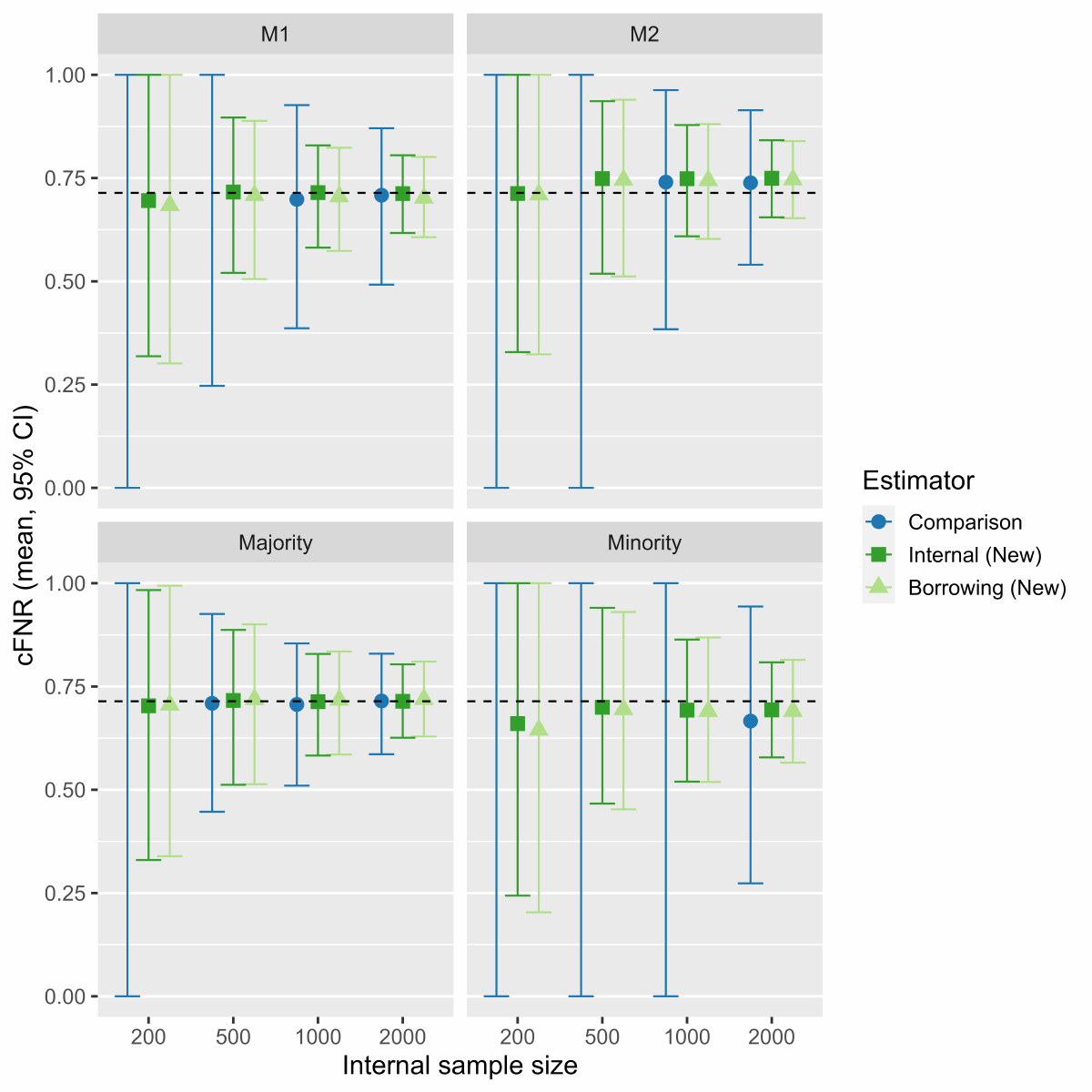}
    \caption{Estimation of $cFNR(S,\pmb{a})$ with varying internal data sample size. Shapes show the mean of $500$ replications and error bars show $95\%$-tile intervals. Dotted horizontal lines show the true value as obtained using the validation set. Error bars without a mean (shape) indicate insufficient data for estimation in at least one replication. In contrast to the comparison estimators, our new estimators achieve reductions in bias and variance and are usable at sample sizes where the comparison estimators cannot be calculated.}
    \label{fig:sim_part1}
\end{figure}

Figure \ref{fig:sim_part1} shows notable variance reduction with our new estimators across sample sizes. Error bars without a mean (shape) indicate the presence of NAs in the replications, i.e. insufficient data for estimation. This occurs with our comparison estimators in all groups at $N_{int}=200$ and in non-majority groups at higher $N_{int}$. In contrast, both our proposed internal and borrowing estimators are able to be calculated for all groups at all sample sizes. Comparing internal and borrowing shows little gain with data borrowing under these particular scenarios.

\subsection{Data borrowing adapts to external data agreement} \label{sec:sims-part2}

This section investigates our data borrowing procedure's ability to respond to varying levels of disagreement in internal and external data distributions. In this section we fix $N_{int}=1000$ and manipulate the level of agreement between internal and external data by multiplying the external data coefficients for $P(\pmb{A}=\pmb{a}|X)$ by a constant in $[-1,1]$.

\begin{figure}
    \centering
    \begin{subfigure}[b]{0.9\textwidth}
         \centering
         \includegraphics[width=\textwidth]{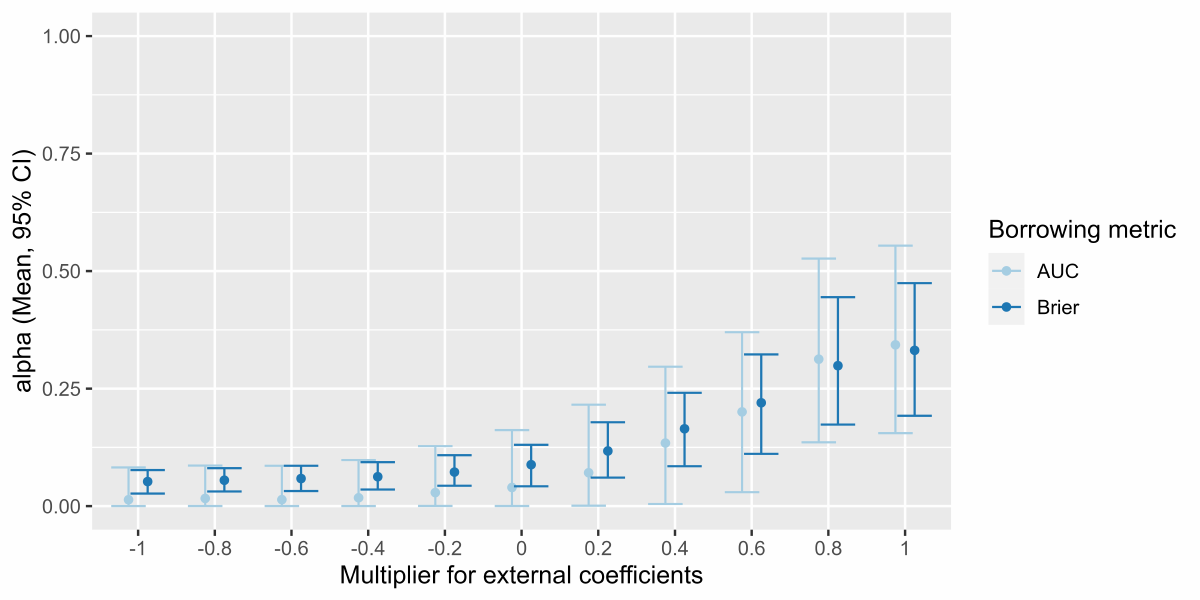}
         \caption{$\alpha$ values at varying levels of agreement between external and internal data. As the data distributions diverge (left side), less information is borrowed from the external data ($\alpha$ approaches zero).}
         \label{fig:sim_part2_a}
     \end{subfigure}
     \hfill
     \begin{subfigure}[b]{.9\textwidth}
         \centering
         \includegraphics[width=\textwidth]{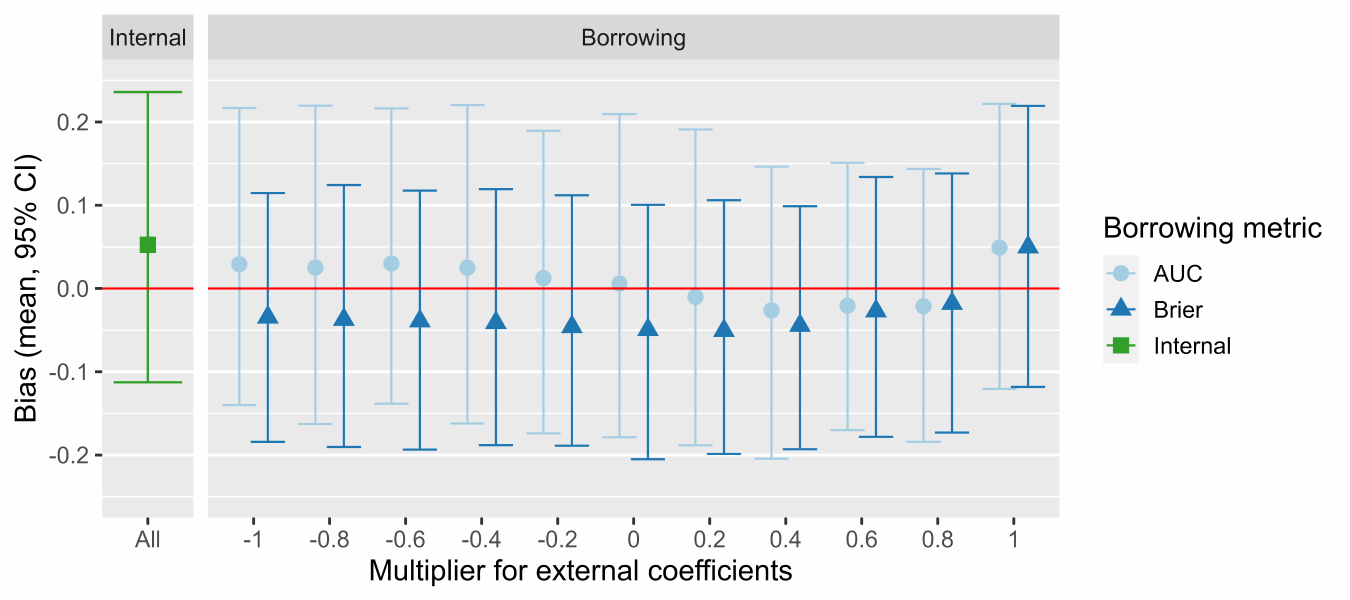}
         \caption{Bias in estimation of minority group $cFNR$ at varying levels of data agreement. Right facet shows bias for internal data estimation (no borrowing).}
         \label{fig:sim_part2_b}
     \end{subfigure}
     \caption{Results of data borrowing with varying levels of agreement between internal and external data distributions. The x-axis shows the factor by which internal data coefficients for generating $P(\pmb{A}=\pmb{a}|X)$ were multiplied to obtain the external data coefficients. Shapes show the mean of 500 replications of the estimation procedure, error bars show 95\%-tile intervals, and colors denote data borrowing metric.}
     \label{fig:sim_part2}
\end{figure}

Figure \ref{fig:sim_part2_a} shows that $\alpha$ decreases as $b$ moves away from $1$. Recall that a lower $\alpha$ means less weight on the external data predictions, i.e., less borrowing. Figure \ref{fig:sim_part2_b} focuses on the minority group $cFNR$ to show that across all distributions, our data borrowing estimators have bias that is less than or equal to the internal data estimation estimator and reasonable variance.

\subsection{Benefit of data borrowing under complex scenarios} \label{sec:sims-part3}

Next we examine scenarios where data borrowing provides gains in bias and variance reduction compared to the internal data version of our estimators. Such gains occur when the internal $h(X, \pmb{a})$ model is difficult to estimate. We use a more complex data generation scenario with an increasing number of noise variables added to the $10$ informative components of $X$ before generating group membership, $P(\pmb{A} = \pmb{a}|X)$, the $Y^0$ potential outcome, $P(Y^0=1|X, \pmb{A})$, and the treatment, $P(D=1 | X, \pmb{A}, S)$. We consider these scenarios with and without 2- and 3-way interactions among four of the $X$ used to generate these probabilities. For simplicity, we focus on the minority group and Brier score for the data borrowing metric. The internal sample size is $500$ throughout.

\begin{figure}
    \centering
    \includegraphics[width=.9\textwidth]{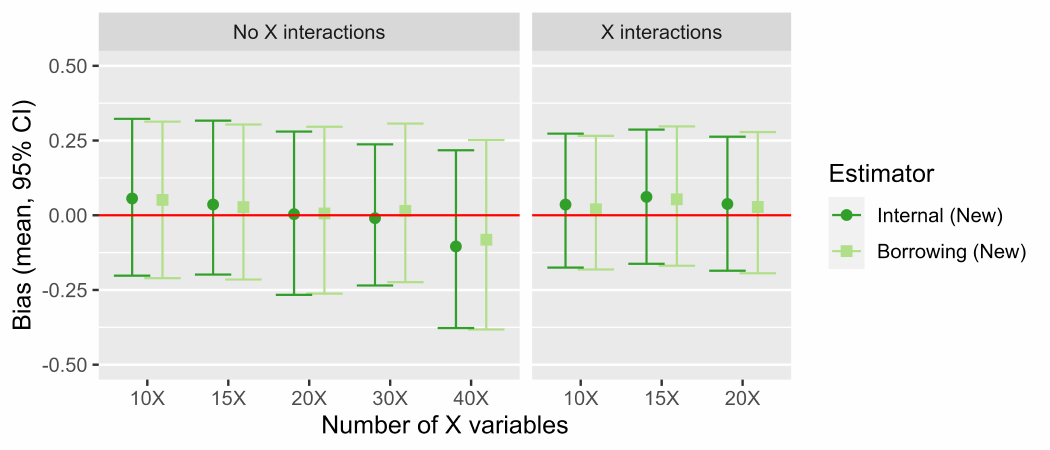}
    \caption{Bias in estimation of minority group $cFNR$ under increasing noise in $X$ with interactions among four of the $X$ (right) and no interactions (left). Shapes show the mean of 500 replications of the estimation, and error bars show 95\%-tile intervals. With no $X$ interactions, bias under data borrowing is similar to internal-only estimation until 40 $X$, where borrowing performs slightly better. With $X$ interactions, bias under data borrowing is similar to internal-only estimation under all scenarios tested. For $\alpha$ values showing how much information is borrowed in each scenario, see Web Appendix D.}
    \label{fig:sim_part3}
\end{figure}

Figure \ref{fig:sim_part3} shows that both with and without $X$ interactions, bias is low and consistent across increasing $X$ noise up to 20 $X$ components. Data borrowing does not notably affect bias except a small reduction under the no-interaction, 40 $X$ components scenario.

\section{COVID-19 risk prediction fairness} \label{sec:application}

During the height of the COVID-19 pandemic, risk prediction models were commonly used to help health systems allocate scarce resources. In addition to their potential benefits, these models raised the challenge of ensuring model performance, and thus resource allocation, was equitable. We studied one such model at a major Midwestern health system that was used to help determine whether a patient should be transferred to a COVID-19 cohort hospital. In our previous work, we analyzed this model's counterfactual error rates across the intersection of age and race, grouping patient self-reported race into white, Black or African American, and all other races. As we noted in that paper, grouping all patients who identify as neither white nor Black/African American together severely limits the utility of the analysis. Here we show how our proposed estimators allow us to disaggregate race and use an additional ethnicity variable to gain a more nuanced picture of model performance. We focus on a single protected characteristic to demonstrate the wide applicability of our proposed methods; our estimators can also be used with multiple, intersecting characteristics.

The risk prediction model was trained on 1,469 adult patients with confirmed or symptomatic suspected COVID-19. Our "internal" data for evaluating the model consisted of 3,649 adult SARS-CoV-2 patients from the same health system who tested positive between 10/27/2020 and 1/9/2022. Deploying our new estimators allowed us to group patient self-reported race and ethnicity variables into five categories: Hispanic or Latino (6.3\%); and non-Hispanic/Latino American Indian or Alaska Native (1.5\%), Asian (4.2\%), Black or African American (13.9\%), and white (74.1\%). We removed patients who marked two or more races, "other" or "all other races" since these records lacked sufficient information to group race/ethnicity. We chose this schema to balance disaggregation with a large enough sample size to obtain useful estimates. Our race categories align with U.S. Census Bureau categories with one exception: we did not have sufficient sample size to break out the Native Hawaiian or Other Pacific Islander category since it had 14 total observations, so we grouped these patients with the Asian category. Likewise, cross-cutting the data by both race and Hispanic or Latino ethnicity was not possible given our sample size, so we opted to consider Hispanic or Latino ethnicity by disaggregating it from the racial groups. We note that the data used in this application does offer relatively rich detail on patients' self-reported race and ethnicity, with 8 and 48 unique responses to the race and ethnicity questions, respectively. Future applications using data sources like this with larger samples could take fuller advantage of the available detail by considering race and ethnicity separately and with more categories.

We chose a cutoff of $0.15$, approximately the $80^{th}$ percentile, for dichotomizing the risk score and used 30-day inpatient readmission or mortality as our outcome and transfer to the cohort hospital as our treatment variable. Covariates comprised comorbidities, home medications, number of prior emergency department visits, and labs and vitals. We excluded covariates with greater than 2/3 missing and performed random forest imputation (missForest package). We considered the same set of covariates for propensity score, outcome, and $h(X,\pmb{a})$ modeling, performing lasso selection separately for each model to select the most relevant variables. As in our prior work, we used a logistic regression propensity score model. Our outcome models were also logistic regressions, and the internal $h(X,\pmb{a})$ model was a single-layer neural network with 50 units (package nnet).

Our external data comprised 8,449 patients from the same health system without a matching SARS-CoV-2 positive test and risk model score. External data patients were grouped into the same five race/ethnicity categories: Hispanic or Latino (8.6\%); and non-Hispanic/Latino American Indian or Alaska Native (1\%), Asian (11.2\%), Black or African American (20\%), and white (59.2\%). While not all internal data covariates were available in the external data, covariates comprised the same general categories. We excluded covariates with greater than 2/3 missing and performed random forest imputation. We performed lasso selection to choose the most important covariates for the external $h(X,\pmb{a})$ model and then fit a single-layer neural network with 100 units using the selected covariates. We used the Brier score data borrowing metric which gave $\alpha = 0.004$. We obtained standard errors using the rescaled bootstrap method proposed in \citet{wastvedtIntersectionalFrameworkCounterfactual2023} and confidence intervals using t-intervals truncated at $0$ and $1$. All analyses were done in R (version 4.2.3, R Core Team 2023).

\begin{figure}
    \centering
    \includegraphics[width=.9\textwidth]{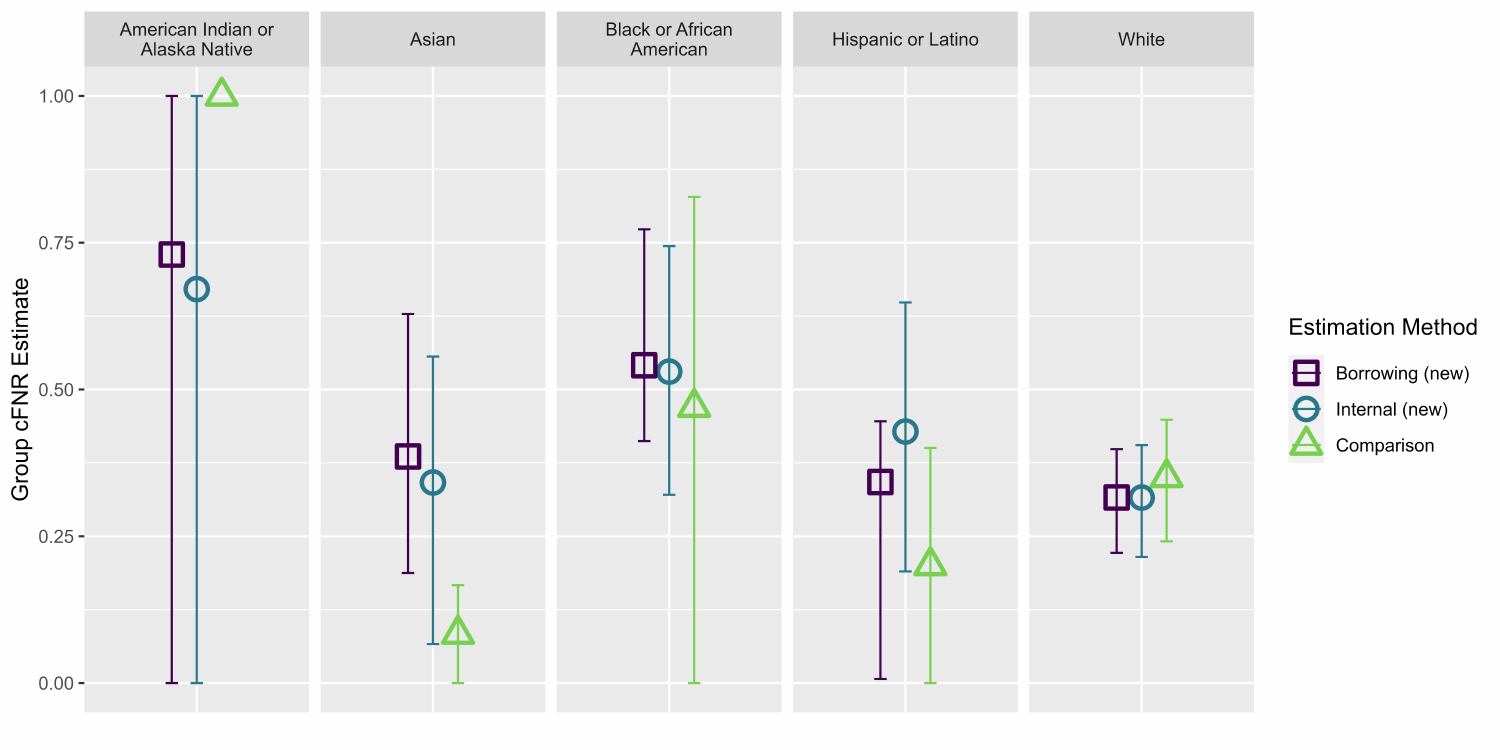}
    \caption{Estimation of group $cFNR$ for 5 racial and ethnic groups using our proposed estimators (internal and borrowing) and our comparison estimator. Shapes show the mean of 200 bootstrap replications, and error bars show 95\% bootstrap t-intervals. Our proposed estimators reduce variance for the Black or African American group and enable estimation for the American Indian or Alaska Native group, although the confidence interval for this group covers the entire $(0,1)$ range. Comparison estimators for Asian, Black or African American, and Hispanic or Latino groups are truncated at $0$. New estimators (internal and data borrowing) for the American Indian or Alaska Native group are truncated at $1$.}
    \label{fig:app_re}
\end{figure}

Our proposed estimators enable estimation for small subgroups where estimation failed using the comparison estimators. Figure \ref{fig:app_re} shows that internal and borrowing estimators reduced variance substantially for the Black or African American group and slightly for the white group. The smallest group, American Indian or Alaska Native, was too small to obtain a confidence interval and obtained the unlikely point estimate of $1$ using the comparison method. Our new estimators obtain an interval and a more reasonable point estimate, although the interval covers the entire $(0,1)$ range and thus has minimal utility. This wide interval demonstrates the sample size limitations that still occur using our proposed estimators.

The Asian and Hispanic or Latino groups had point estimates that differ substantially under the new and comparison methods, which may be due to undercoverage issues for the comparison method. Estimates for these groups using the comparison method are also highly variable because of extremely small counts ($<10$) in the false negative cell of the confusion matrix, which is the relevant cell for the comparison $cFNR$ estimate. Note also that while the confidence interval for the Asian group appears smaller under the comparison method, this is due to the truncation of intervals at zero. Intervals affected by truncation are noted in the caption to Figure \ref{fig:app_re}. 

\section{Discussion} \label{sec:discussion}

In this paper we took a three-pronged approach to addressing the common challenge of small subgroups in risk prediction model fairness assessments. First, we proposed new estimands that reformulate the counterfactual error rates to borrow strength across groups. We showed how these estimands can be identified from observed data in a manner that draws on more of the data than existing alternatives. Finally, we proposed a novel data borrowing method with the potential to leverage external data to further improve estimation.

Our methods have limitations: most importantly, subgroups can still be too small for the methods to produce useful estimates. When sample size is too small, confidence intervals may be too wide, as in the American Indian or Alaska Native group in our application (Section \ref{sec:application}), or estimates may differ substantially between our proposed method and the comparison method. However, this sample size cutoff is fairly small, as demonstrated by the useful estimates produced for the Black or African American group in our application, a group which only had $11$ observations in the false negative cell of the confusion matrix (see Web Table 1 for confusion matrices by group). We refrain from offering a specific sample size cut-off for our methods since this number will depend on the richness of the covariates available in a given application, both for model fitting and for meeting our additional assumption described in Section \ref{sec:estimands}. Our proposed estimators also require practitioners to fit more models than existing alternatives which is both more time-consuming and provides more opportunities for model mis-specification. Nevertheless, we find little impact of model mis-specification under the scenarios considered in our simulations (Web Appendix D).

Finally, our estimators allow practitioners to look at more disaggregated protected groups, which we hope will facilitate more nuanced consideration in fairness assessments of groups that are both marginalized and small. In the case of race and ethnicity, as in our COVID-19 application, our estimators are an improvement over those that require grouping disparate subgroups or excluding small groups altogether. However, given the prevalence of using racial categories as a protected characteristic in this field, it is important to recognize the limitations inherent in any race-based fairness assessment. No matter how disaggregated, racial categories are not intrinsic identities; rather, they are socially constructed labels that comprise "a system of inherently unequal status categories" \citep{benthallRacialCategoriesMachine2019}. In a critique of traditional approaches to algorithmic fairness, \citet{weinbergRethinkingFairnessInterdisciplinary2022} frames the problem as one of abstraction: by treating race as an intrinsic quality, researchers minimize the structural factors that create and maintain racial inequality. Whether conducted with our proposed metrics or others, a fairness assessment involving race bears the responsibility to drive toward undermining the systems of hierarchy and power imbalances that cause algorithmic unfairness. \citet{hannaCriticalRaceMethodology2020} suggest considering descriptive analyses of model performance across the many dimensions of race such as self-identified, other-identified, phenotype, and others. Differing patterns of unfairness can provide more insight into how to proceed to ameliorate inequities. Finally, \citet{hannaCriticalRaceMethodology2020} emphasize the importance of looking beyond the algorithm to promote justice at all parts of the process, including the data models are trained on, the choice of where and on whom to deploy them, and the way model predictions are incorporated into human practices.  

\section{Figures and Tables}

\FloatBarrier


\backmatter

\section*{Acknowledgements}

This work was supported by the University of Minnesota Doctoral Dissertation Fellowship. Data access was facilitated by the University of Minnesota's Clinical Quality, Outcomes, Discovery, and Evaluation Core (CQODE).



\bibliographystyle{biom} \bibliography{paper3}

\label{lastpage}

\end{document}